\documentclass[a4paper,12pt]{article}
\usepackage{jcappub} 
\usepackage{lineno}

\usepackage{amsmath,amssymb,graphicx,subfigure,dcolumn,multirow,bm,soul,xcolor}

\definecolor{DarkBlue}{rgb}{0.15,0.15,0.85}
\usepackage[linktocpage=true]{hyperref}
\graphicspath{{Figures/}}
\DeclareMathOperator{\sech}{sech}
\DeclareMathOperator{\csch}{csch}

\title{\boldmath Development of generic no-scale inflation}

\author[a]{Lina Wu,}
\emailAdd{wulina@xatu.edu.cn}
\author[a]{Jin-Ke Shen,}
\emailAdd{q1668179812@163.com}
\author[b,c]{Tianjun Li,}
\emailAdd{tli@itp.ac.cn}
\author[,d,e]{and Junle Pei \footnote{Corresponding author.}}

\emailAdd{peijunle@ihep.ac.cn}

\affiliation[a]{School of Sciences, Xi'an Technological University, Xi'an 710021, P. R. China}
\affiliation[b]{CAS Key Laboratory of Theoretical Physics, Institute of Theoretical Physics, Chinese Academy of Sciences, Beijing 100190, P. R. China}
\affiliation[c]{School of Physical Sciences, University of Chinese Academy of Sciences, No.~19A Yuquan Road, Beijing 100049, P. R. China}
\affiliation[d]{Institute of High Energy Physics, Chinese Academy of Sciences, Beijing 100049, China}
\affiliation[e]{Spallation Neutron Source Science Center, Dongguan 523803, China}

\abstract{ We develop generalized no-scale supergravity models of inflation, and then study the corresponding cosmological predictions as well as the formation of primordial black holes (PBHs) and scalar-induced gravitational waves (SIGWs).
With a new parameter $0<a\leq 1$, the generalized no-scale supergravity provides the continuous connections among the generic no-scale supergravity from string theory compactifications.
The resulting prediction of the CMB, spectrum index $n_s$, and tensor-to-scalar ratio $r$ can be highly
consistent with the latest Planck/BICEP/Keck Array observations. Notably, the models with $a\neq 1$ give a smaller ratio $r\leq 10^{-3}$, which is flexible even under the anticipated tighter observational constraints at the future experiments.  Additionally,  these models have the potential to generate a broad-band stochastic gravitational wave background, and thus explain the NANOGrav 15yr signal. Furthermore, they predict the formation of PBHs with various mass scales, which could account for a significant portion of dark matter relic density in the Universe.}

\begin{document}
\maketitle
\flushbottom

\section{Introduction}

The early Universe is believed to have gone through a prolonged period of acceleration, termed as the cosmic inflation \cite{Starobinsky:1980te,Guth:1980zm,Linde:1981mu}, to solve the problems of the big bang theory, such as the horizon problem, flatness problem, and other related problems. Additionally, inflation can amplify the primordial density (curvature) perturbations arising from quantum fluctuations, thereby explaining the measured temperature fluctuations of cosmic microwave background (CMB) radiation and providing a potential origin of seeds for the subsequent formation of structures. A significant number of single field inflationary models have been developed and refined through temperature and polarization measurements on the CMB anisotropy. Combining with the BICEP/Keck data and Planck 2018 results, the scalar spectral index $n_s$, tensor-to-scalar ratio $r$, and scalar amplitude $A_s$ for the power spectrum of the curvature perturbation are currently constrained to be $n_s=0.9649\pm 0.0042$, $r_{0.05}\leq 0.032$ (95\%~C.L.), and $A_s=2.10\times10^{-9}$, respectively \cite{Planck:2018jri,BICEP:2021xfz,Tristram:2021tvh}. Furthermore, future experiments are expected to impose even stricter constraint on tensor-to-scalar ratio approaching $r\sim 10^{-3}$ \cite{LiteBIRD:2022cnt}. It is therefore imperative to find the classes of models that satisfy such stringent constraint.

A few kinds of inflationary models, such as the $R+R^2$ Starobinsky model \cite{Ellis:2013xoa,Ellis:2021kad,Ellis:2013nxa,DeFelice:2023psw,Jeong:2023zrv,Brinkmann:2023eph}, the $\alpha$-attractor E/T models \cite{Kallosh:2013yoa,Gao:2017uja,Sabir:2019wel,Ellis:2019bmm,Ellis:2020xmk,Iacconi:2021ltm}, and the Higgs inflation with non-minimal coupling~\cite{Kawai:2014gqa,Chen:2018ucf,Karydas:2021wmx,Shaposhnikov:2020gts} have the attractive feature that the predicted scalar spectral tilt $n_s\simeq 1-2/N$ and the predicted tensor-to-scalar ratio $r\simeq 12/N^2$ with \textit{e}-folds $N=50-60$ are in compliance with the CMB observations. The $N$ indicates the number of \textit{e}-folds between the horizon exit of CMB modes and the end of inflation. Moreover, the whole observational $n_s-r$ plane posted by the Planck/BICEP/Keck data is covered by the complement of ``exponential $\alpha$-attractor" models and the ``polynomial  $\alpha$-attractor" models \cite{Kallosh:2022feu,Bhattacharya:2022akq}. The exponential models are on the left with smaller $n_s$, while the polynomial models are on the right with larger $n_s$. A bayesian analysis of a generalization of $\alpha$-attractor T model \cite{German:2021rin} shows that the standard $\alpha$-attractor model with tangent potential is favoured by the current CMB data \cite{LinaresCedeno:2022rbq}.
More surprising to particle physicists is that these models can all be embedded in supergravity scenario \cite{Ellis:2015xna,Salvio:2017xul,Pallis:2023nzl,Ellis:2020lnc}. 

Recently, a class of multi-moduli inflation has been realized in the generic no-scale supergravity inspired by string theory compactifications \cite{Wu:2021zta,Wu:2022kew}, in which the K\"ahler potential is
\begin{equation}
    K=-N_1\log({T_1+\overline{T}_1}-2|\varphi|^2)-\sum_{i=2,3}N_i\log({T_i+\overline{T}_i})~,
\end{equation}
where $T_i$ are the K\"ahler moduli, and $\varphi$ denotes the matter, Higgs, and inflaton fields. In addition, $N_i$ are positive integers, and satisfy the equation $N_1+N_2+N_3=3$ as required by the no-scale supergravity. The spectral index is $n_s\simeq 1-2/N$ for these models with one, two, and three moduli. The predicted tensor-to-scalar ratio is $r\simeq a_i/N^2$ for the models with one and three moduli ({$a_1=12$ and $a_3=4$}), while $r\simeq 84/N^4$ for the model with two moduli. These predictions are well compatible with the current and future CMB observations.  The attractor T and E models have been reconstructed. Additionally, the quadratic and quartic inflationary models embedded in generic no-scale supergravity exhibit a plateau for the inflaton field and are still able to survive under the stringent constraints on $r$ \cite{Wu:2022kew}. It provides a viable framework to explain the inflationary epoch of the early Universe while consistent with the observational data.

In this paper, we study the generic no-scale supergravity whose  K\"ahler potential is 
\begin{align}
    K&=-3a\ln{(T+\overline{T}- |\varphi|^2)}-3(1-a)\ln{(T'+\overline{T'})}~.
\end{align}
{The kinetic term for the inflaton field $\varphi$ must be positive, whereas the kinetic term for the modulus $T'$ should be non-negative. Consequently, the parameter $a$ is constrained to the range of $0< a\leq 1$. This ensures that this model continuously connect the above three models to each other.}
When $a$ is equal to $1$, $2/3$, and $1/3$, it gives us the no-scale supergravity with one, two, and three complex moduli from string theory compactifications, respectively.

With the Wess-Zumino superpotential, we consider the inflationary model in the generic no-scale supergravity in Section \ref{sect:setup},
and study the cosmological predictions $n_s$ and $r$ for various $a$ regimes in Sections \ref{sect:inf_per} and \ref{sect:ana}. Accompanied by the numerical results and plots, we also try to give an analytic understanding on the evolution of the slow-roll parameters $\varepsilon$ and $\eta$, where a hierarchy occurs between the magnitude of these two parameters at the horizon exit. Thus, $r$ can be smaller than $10^{-3}$ while $n_s$ can locate in the $1\sigma$ regime of Planck/BICEP/Keck data. In Section \ref{sect:pbh}, we show the benchmark points for the models where the power spectra are enhanced to be $\mathcal{O}(10^{-2})$ at small scale. The results show that the generated PBHs can account for almost all of dark matter relic density, and the induced gravitational waves (GW) can be a source of stochastic GW background given by NANOGrav 15yr data. Finally in Section \ref{sect:con}, we present the conclusion with a brief discussion. 
 
\section{Inflationary Potential from Supergravity}\label{sect:setup}
The Lagrangian with the complex scalar field $\Phi^i$ for $\mathcal{N}=1$ supergravity can be written as
\begin{equation}
    \mathcal{L}\propto \sqrt{-g}\left(K_i^{\bar j}\partial_{\mu}\Phi^i\partial^{\mu}\bar{\Phi}_{\bar j}-V(\Phi)\right)~,~
\end{equation}
where the K\"ahler metric is defined as
$  K_i^{\bar j}=\partial^2 K/\partial \Phi ^i\partial \bar{\Phi}_{\bar j} $.
The structure of the supergravity is characterized by the K\"ahler potential $K$ and superpotential $W$, and the effective scalar potential can be written in the form of
\begin{equation}
V=e^K\left[D_iW\left(K^{-1}\right)^i_{\overline{j}}D^{\overline{j}}\overline{W}-3|W|^2\right]~,~\label{eq:pot-sugra}
\end{equation}
where the K\"ahler covariant derivative is $D_i W\equiv W_i +K_iW$.
The K\"ahler potential and superpotential are given in the form of 
\begin{align}
	K&=-3a\ln{(T+\overline{T}- |\varphi|^2)}-3(1-a)\ln{(T'+\overline{T'})}~,\\
	W&=\frac{M }{2}\varphi ^2-\frac{\lambda }{3}\varphi ^3~,
\end{align}
where the parameter $0<a\leq1$. When $a$ is setting as $1$, $2/3$ or $1/3$, the models become no-scale supergravity with one, two, or three complex moduli, respectively. And the details can be found in Refs.~\cite{Wu:2021zta,Wu:2022kew}. 
Following the stabilization of the moduli fields as $<T>=c_1/2$ and $<T'>=c_2/2$ \cite{Ellis:2013xoa,Wu:2016fzp,Ellis:2018ojk,Wu:2021zta,Wu:2022kew}, and assuming that the inflation goes along the real components of the matter field $\varphi$, we can get the scalar potential in the Jordan frame as
\begin{equation}
	V_J(\varphi)=V_0  \varphi ^2 (1-\beta \varphi)^2\left(c_1-\varphi^2\right)^{1-3 a }~,~
\end{equation}
where $V_0= M^2c_2^{3(a-1)}/3a$ and $\beta= \lambda/M$.
	
Note that the inflaton is noncanonical, defining a new canonical field $\chi=(x+\mathrm{i} y)/\sqrt{2}$, we consider the field transformation
\begin{equation*}
	\partial_{\mu}\chi \partial^{\mu}\chi=K_{\varphi\overline{\varphi}}\partial_{\mu}\varphi\partial^{\mu}\overline{\varphi}
\end{equation*}
with
\begin{equation}\label{eq:fieldtrans}
	K_{\varphi\bar{\varphi}}=\frac{3 a  (T+\overline{T})}{\left(T+\overline{T}-\varphi ^2\right)^2}~.
\end{equation}
Then the corresponding complex fields in Einstein and Jordan frame are respectively
\begin{equation*}
	\begin{split}
		\chi=\sqrt{6a}  \tanh ^{-1}\left(\frac{\varphi }{\sqrt{c_1}}\right)~,~~
		\varphi=\sqrt{c_1}\tanh{\left(\frac{\chi}{\sqrt{6a}}\right)}~.
	\end{split}
\end{equation*}

In terms of the real part of the new inflaton field $\chi$, the inflaton potential in the
Einstein frame is
\begin{equation}
	V_E(x) = V_0 \tanh ^2\left(\frac{x }{\sqrt{6a}  }\right) \text{sech}^2\left(\frac{x }{\sqrt{6a } }\right)^{1-3 a } \left(1-\beta \tanh \left(\frac{x }{\sqrt{6a}  }\right) \right)^2~. \label{eq:pot-gen}
\end{equation}
Without loss of generality, we will only consider the inflationary models with positive $\beta$.

\section{Inflation and Cosmological Perturbations}\label{sect:inf_per}

The well-known slow-roll parameters defined by the scalar potential are given by
\begin{equation}
    \varepsilon=\frac{M_{\rm Pl}^2}{2}\left(\frac{V_E'}{V_E}\right)^2~,~~\eta=M_{\rm Pl}^2\frac{V_E''}{V_E}~. \label{eq:sr_para}
\end{equation}
In the slow-roll approximation ($\varepsilon\ll 1,~\eta \ll 1$) and neglecting the contribution from the higher order slow-roll corrections \cite{Li:2014zfa}, the CMB observations including the scalar spectral index, the tensor-to-scalar ratio, and the amplitude of the power spectrum are predicted as 
\begin{equation}
    n_s=1-6\varepsilon(x_0)+2\eta(x_0)~,~~ r=16\varepsilon(x_0)~,~~ A_s=\frac{1}{24\pi^2M_{\rm Pl}^2}\frac{V_E(x_0)}{\varepsilon(x_0)}~, \label{eq:CMB}
\end{equation}
and the \textit{e}-folding number is 
\begin{equation}
    N=-\frac{1}{M_{\rm Pl}^2}\int_{x_0}^{x_e}{\rm d}x \frac{V_E}{V_E'}~,
\end{equation}
where $x_0$ and $x_e$ are the value of field when the interesting mode $k_*$ crossed outside the horizon and the value of field at the end of inflation. In the paradise of slow-roll inflation, the expansion ends when $\varepsilon=1$ or $\eta=1$. In our calculations, the amplitude of the power spectrum is fixed to the central value $A_s=2.10\times 10^{-9}$ \cite{Planck:2018jri,BICEP:2021xfz} by formulating the parameter $V_0$. In the following sections, where not specified, $V$ will indicate the scalar potential $V_E$ in Einstein frame and we will use the units in which the reduced Planck mass is set to 1. 

The Hubble constant is $H=\dot{a}/a$ and thus the Hubble slow-roll parameters are defined as \cite{Baumann:2022mni}
\begin{equation}
   \varepsilon_H=-\frac{\dot{H}}{H^2}~,~~\eta_H=\varepsilon_H-\frac{1}{2H}\frac{\dot{\varepsilon}_H}{\varepsilon_H}~. 
\end{equation} 
The equation of motion for the inflaton fluctuation $v=a\delta x$, namely Mukhanov-Sasaki equation \cite{Sasaki:1986hm,Mukhanov:1988jd}, is 
\begin{equation}
    \frac{{\rm d^2} v_k}{{\rm d}\tau^2}+\left(k^2-\frac{1}{z}\frac{{\rm d^2}z}{{\rm d}\tau^2}\right)v_k^2=0~, \label{eq:muk_sas}
\end{equation}
where the conformal time ${\rm d}\tau=a {\rm d} t$, and the parameter $z=a\sqrt{2\varepsilon_H}$. 
In the limit $\tau\to-\infty$, the mode of interest is well inside the Hubble radius, i.e., the initial condition for the mode $v_k$ is the Bunch-Davis vacuum,
\begin{equation}
    v_k|_{k\tau\to-\infty}=\frac{1}{\sqrt{2k}}e^{-ik\tau}~.
\end{equation}
At the end of inflation, the inflaton ceases to fluctuate, and subsequently, the scalar curvature perturbation translates into the primordial density fluctuation of the hot Big Bang \cite{Baumann:2018vus}. The scalar mode can be rewritten in terms of the comoving curvature perturbation $\mathcal{R}$ as 
\begin{equation}
    v_k=-z\mathcal{R}_k~,
\end{equation}
and the power spectrum is defined by
\begin{equation}
    \mathcal{P_R}=\frac{k^3}{2\pi^2}\left|\frac{v_k}{z}\right|^2_{k\ll aH}~.
\end{equation}
Within the slow-roll approximation, the CMB observations can be obtained by analytically solving  Eq.~\eqref{eq:muk_sas} and shown in Eq. \eqref{eq:CMB}. While the inflation proceeds ultra-slow-roll phase \cite{Tsamis:2003px,Kinney:2005vj}, where the potential is perfectly flat and the slow-roll is violated, we should numerically solve the Eq.~\eqref{eq:muk_sas}, in which the derivative with respect to $\tau$ converts to derivative with respect to $N$~\cite{Lin:2020goi,Yi:2020cut,Gao:2020tsa,Wu:2021zta,Lin:2021vwc,Cai:2023uhc,Wang:2024euw}. The section \ref{sect:pbh} reveals that inflation undergoes an ultra-slow-roll phase where PBHs and induced GWs are generated as it approaches the inflection point.

The Friedman equation and the Klein-Gordon function  can be written as
\begin{equation}
    \begin{split}
        x''+(3-\varepsilon_H)x'=0~,~~
        H^2=\frac{2V}{6-x'^2}~.
    \end{split}
\end{equation}
The prime stands for derivative with respect to $N=\ln{a}$. During the slow-roll states, the Hubble parameter $\varepsilon_H$ can be neglected. Integrating the above function, we can get the velocity of the inflaton,
\begin{equation}
    x'=e^{-3N}~.
\end{equation}
Then the Hubble slow-roll parameters become
\begin{equation}
    \varepsilon_H=\frac{1}{2}x'^2~,~~\eta_H=\varepsilon_H-\frac{1}{2}\frac{\varepsilon_H'}{\varepsilon_H}~. 
\end{equation}
When inflation proceeds the ultra-slow-roll phase, the first parameter is $\varepsilon_H\simeq\varepsilon_{SR}\sim 0$, and the second parameter becomes $\eta_H\simeq 3$, which violates the slow-roll conditions. The curvature power spectrum becomes 
\begin{equation}
    \mathcal{P_R}=\frac{H^2}{8\pi^2\varepsilon_H}\sim e^{6N}~.
\end{equation}
This implies that whenever the potential exhibits a perfectly flat region, such as around an inflection point, the power spectrum experiences a notable enhancement. This enhancement is pivotal for the generation of primordial black holes and induced gravitational waves. Evidence of this can be found in the following papers \cite{Ballesteros:2017fsr,Mishra:2019pzq,Lu:2019sti,Fu:2019ttf,Dalianis:2019vit,Lin:2020goi,Yi:2020cut,Gao:2020tsa,Wu:2021zta,Zhang:2021rqs,Heydari:2021qsr,Rezazadeh:2021clf,Ahmed:2021ucx,Kawai:2021edk,Spanos:2021hpk,Cai:2021wzd,Lin:2021vwc,Pi:2021dft,Pi:2022zxs,Kallosh:2022vha,Kawai:2022emp,Braglia:2022phb,Fu:2022ssq,Fu:2022ypp,Mavromatos:2022yql,Yi:2022anu,Meng:2022low,Qiu:2022klm,Mu:2022dku,Mu:2023wdt,Cai:2023uhc,Poisson:2023tja,Heydari:2023rmq,Ghoshal:2023wri,Zhao:2023xnh,Arya:2023pod,Yi:2023npi,Feng:2023veu,Solbi:2024zhl,Wang:2024euw,Chen:2024gqn,Zhao:2024yzg,Yang:2024ntt,Choudhury:2024jlz,Gao:2024csn} and the references in there.

\section{Inflationary predictions}\label{sect:ana}
When setting $a=1$ and $\beta=1$, the potential in Eq. \eqref{eq:pot-gen} becomes $V=\frac{V_0}{4}\left(1-e^{-\sqrt{\frac{2}{3}} x }\right)^2$. That is to say that the Starobinsky inflation model or standard E-model inflation is realized \cite{Ellis:2013xoa}.  This model has been widely discussed previously  in the supergravity scenario~\cite{Ellis:2013xoa,Wu:2022kew}, and its cosmological predictions fit the current experimental CMB data very well. In the following, we will discuss the predictions of the  inflationary potential with $a\neq 1$.

\subsection{$0<a<1/3$}	
\begin{figure}[th]
    \centering
    \subfigure[]{\includegraphics[height=0.30\linewidth]{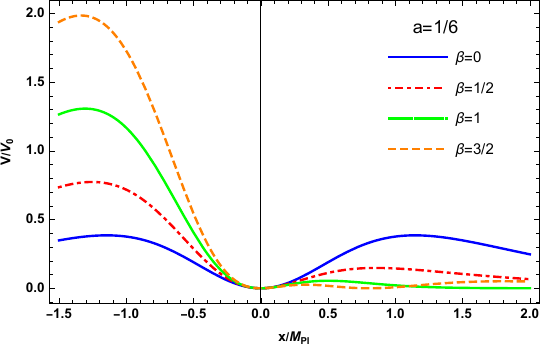}}\quad
    \subfigure[]{\includegraphics[height=0.30\linewidth]{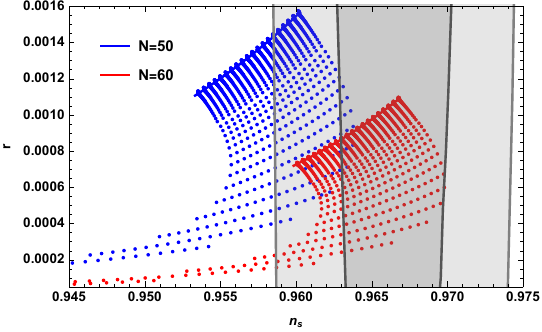}}
    \caption{(a) {Potential with $a=1/6$. }There are three and five extreme points for $\beta\leq1$ and $\beta>1$. (b) The observations $n_s$ \textit{vs} $r$ for the models with parameters $a\lesssim1/3$.}
    \label{fig:pot_16}
\end{figure}

There are three and five extreme points for the potential with $\beta\leq1$ and $\beta>1$ (see Fig. \ref{fig:pot_16}(a)), and the extreme points are located at $x_{M_i}$ (maximum) and $x_{m_i}$ (minimum). The subscript $i$ increases from left to right and $x_{m_1}=0$. If $0<a<1/3$, then $1-3a>0$. In this case, the potential in Eq. \eqref{eq:pot-gen}  becomes $V(x)\to 0$ in the limit $x\to \pm\infty$. In order to get proper $n_s$ and $r$, $a$ should be very closed to $1/3$.
The slow roll parameters are approximate as
\begin{eqnarray}
	\varepsilon(x)&=&F(x)\left[\left(\frac{1}{3a}+3a-2\right)+2G(x)
	\left(\frac{1}{3a}-1\right)+...\right]~,\\
	\eta(x)&=& F(x)\left[2\left(\frac{1}{3a}+3a-2\right)+G(x)\left(\frac{10}{3a}-7\right)+...\right]~,
\end{eqnarray}
where $F(x)=\frac{1}{\left(\beta-\coth\left(\frac{x}{\sqrt{6a}}\right)\right)^2}$, $G(x)=\frac{4\beta}{\sinh\left(\sqrt{\frac{2}{3a}x}\right)}$.
In the limit $x\to \pm\infty$, the slow-roll parameters are 
\begin{equation}
	\varepsilon=3 a +\frac{1}{3 a }-2~, ~~\eta=2\left(3 a +\frac{1}{3 a }-2\right)~.
\end{equation}
Therefore, the trajectories from $x_{M_i}$ to $\pm\infty$ are prohibited, as the expansion time of the early universe is insufficient to align with the current CMB measurements.  Conversely, other inflationary trajectories, originating from $x_M$ and terminating at $x_m$, produce the appropriate spectrum index and tensor-to-scalar ratio, which are consistent with Planck+BICEP data and illustrated in Fig. \ref{fig:pot_16}(b).

\subsubsection{Trajectory I: $x_{M_1}<x_i<x_e<x_{m_1}$}
The first possible inflationary trajectory is from $x_{M}<0$ to $\phi_m=0$, and the allowable parameter range is $a\lesssim 1/3$. 
Without loss of generality, we set $\beta=0$, and the potential becomes
\begin{equation}
    V(x)=\tanh ^2\left(\frac{x}{\sqrt{6a } }\right) \text{sech}^2\left(\frac{x}{\sqrt{6 a }}\right)^{1-3 a }~.
\end{equation}
The slow-roll parameters are given by
\begin{equation}
    \begin{split}
        \varepsilon(x)&=\frac{ 1}{3 a }\tanh ^2\left(\frac{x}{\sqrt{6a} }\right)\left(-1+3 a +\text{csch}^2\left(\frac{x}{\sqrt{6a} }\right)\right)^2 \qquad\qquad\qquad\qquad\quad\\
        &\simeq\frac{4}{3a}\csch^2\left(\sqrt{\frac{2}{3a}}x\right)~,
    \end{split}
\end{equation}
\begin{equation}
    \begin{split}
        \eta(x)&=\frac{1}{3 a }\left(2 (1-3 a )^2+\text{csch}^2\left(\frac{x}{\sqrt{6a}}\right)-(3 a -2) (6 a -5) \text{sech}^2\left(\frac{x}{\sqrt{6a}}\right)\right)\\
        &\simeq \left(5-\frac{7}{3a}\right){\sech}^2\left(\frac{x}{\sqrt{6a}}\right)~.
    \end{split}
\end{equation}
Then,
\begin{equation}
    \begin{split}
        1-n_s(x)
        &\simeq 2\left(1+\frac{1}{3a}\right){\sech}^2\left(\frac{x}{\sqrt{6a}}\right)~.\\
    \end{split}
\end{equation}
The \textit{e}-folding number is 
\begin{equation}
    N(x)=\frac{3 a \log \left(3-3 a +(3 a -1) \cosh \left(\sqrt{\frac{2}{3a}} x\right)\right)}{2 (3 a -1)}~.
\end{equation}
When the slow-roll parameter satisfies $\varepsilon(x)=1$, the inflation ends at
\begin{equation}
    \begin{split}
        x_{e}&=\pm\sqrt{\frac{3a }{2}} \sinh ^{-1}\left(\frac{2}{\sqrt{3a}}\right)~, \\
        N(x_{e})&=\frac{3 a \log \left(3-3 a +(3 a-1) \sqrt{\frac{4}{3 a }+1}\right)}{6 a -2}~.\\
    \end{split}
\end{equation}
{Defining new parameters 
\begin{equation}
    y\equiv\frac{x}{\sqrt{6a}}~,~~ Y\equiv\cosh{2y}~,
\end{equation}
the \textit{e}-folding number and the slow-roll parameters become
\begin{equation}
\begin{split}
    N(x)&=\frac{3 a  \log \left(3-3 a +(3 a -1) Y\right)}{2 (3 a -1)}~,\\
    \varepsilon(Y) &\simeq \frac{4}{3a  \left(Y^2-1\right)}~, \\
    \eta(Y) &\simeq \left(5-\frac{7}{3 a }\right)\frac{2 }{Y+1}~,
\end{split}
\end{equation}
with $$\sech^2{y}=\frac{2}{1+\cosh{2y}}=\frac{2}{1+Y}~,~~\csch^2{y}=\frac{2}{\cosh{2y}-1}=\frac{2}{Y-1}~.$$
The parameter $Y$ can be rewritten in terms of the \textit{e}-folding number,
\begin{equation}
    Y=\frac{3 a +e^{\left(2-\frac{2}{3a }\right) N}-3}{3a -1}\sim 1+\frac{2}{1-3a}~.
\end{equation}
Then the spectrum index and tensor-to-scalar ratio are
\begin{equation}
    \begin{split}
        n_s(Y)&\simeq 1-4\left(1+\frac{1}{3a}\right)\frac{1}{1+Y}~,\\
        r(Y)&\simeq\frac{64}{3 a  \left(Y^2-1\right)}~,
    \end{split}
\end{equation}
or
\begin{equation}
    n_s(N)=1-\frac{4 \left(9 a ^2-1\right)}{3a \left(6 a +e^{\left(2-\frac{2}{3 a }\right) N}-4\right)}~.
\end{equation}
When $Y\sim 10^{2}-10^{3}$, we have $n_s \sim 0.92-0.99$ and $r\sim 10^{-5}-10^{-3}$. The analytical results match well with the numerical results, which cover the Planck/BICEP plane on the left. The plots for the spectrum index \textit{vs} the tensor-to-scalar ratio are shown in Fig.~\ref{fig:nsr_16}. As $a$ decreases, the observation $r$ decreases whereas $n_s$ increases. 

\begin{figure}[t]
    \centering
    \subfigure[]{\includegraphics[height=0.28\linewidth]{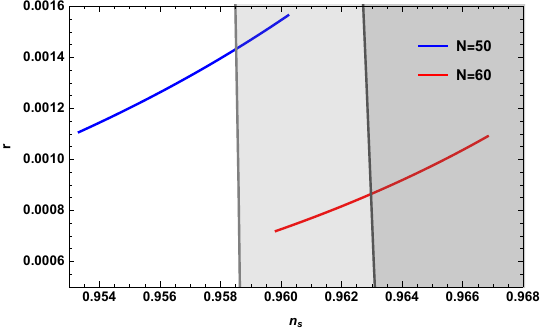}}\quad
    \subfigure[]{\includegraphics[height=0.28\linewidth]{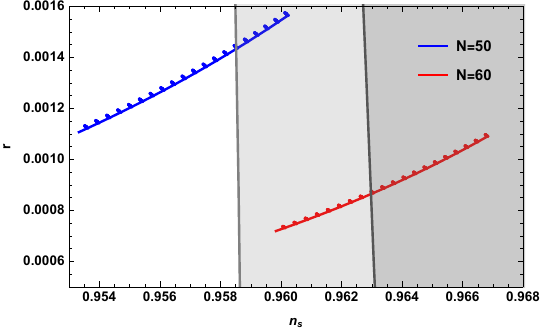}}
    \caption{The observations $n_s$ \textit{vs} $r$ for the models with parameters $a\lesssim1/3$.  The two panels are corresponding to $\beta=0$ (a), $0\leq\beta<10$ (b).}
    \label{fig:nsr_16}
\end{figure}

\begin{figure}[t]
    \centering
    \includegraphics[height=0.26\linewidth]{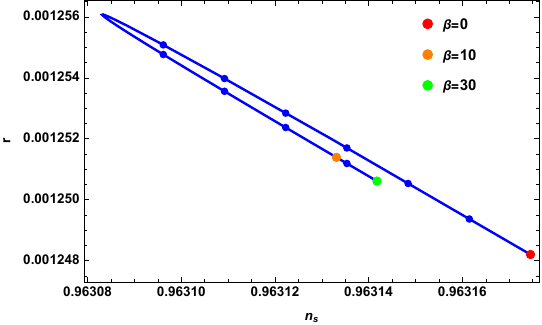}\quad
    \includegraphics[height=0.28\linewidth]{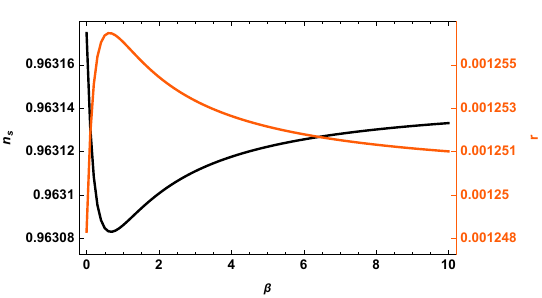}
    \caption{The observations $n_s$ \textit{vs} $r$ locate the $1\sigma$ region of Planck data with fixed $a=1/3.001$ and $N=55$. }
    \label{fig:nsr_16_beta}
\end{figure}

From Fig.~\ref{fig:nsr_16}(b), it is evident that the predictions are not significantly influenced by the parameter $\beta$. Hence, we can initially set $\beta=0$ in this subsection. To delve deeper into the dependency on $\beta$, we have also computed the predictions $n_s$ and $r$ for the models with fixed $a=1/3.001$, as presented in Fig.~\ref{fig:nsr_16_beta}. As $\beta$ increases, the spectrum index $n_s$ exhibits a trend of initially decreasing and then subsequently increasing. Concurrently, the tensor-to-scalar ratio $r$ follows an opposite pattern, first increasing and then decreasing. As $\beta$ continues to rise, both $n_s$ and r tend to stabilize, indicating a convergence towards specific values,
\begin{equation}
    n_s= 0.96314~,~~r= 1.25\times10^{-3}~. 
\end{equation}
}

\subsubsection{Trajectory II: $0=x_{m_1}<x_e<x_i<x_{M_2}$}

\begin{figure}[th]
    \centering
    \subfigure[]{\includegraphics[height=0.28\linewidth]{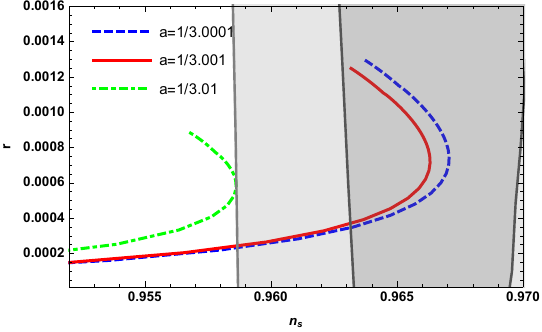}}\quad
    \subfigure[]{\includegraphics[height=0.28\linewidth]{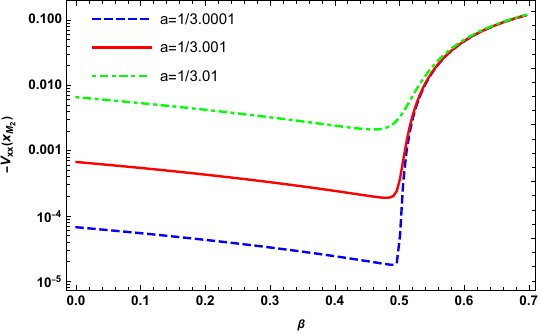}}
    \subfigure[]{\includegraphics[height=0.28\linewidth]{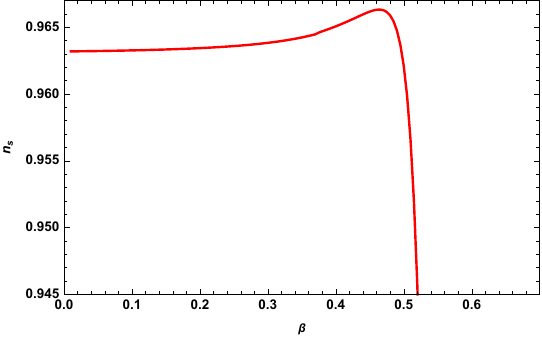}}\quad
    \subfigure[]{\includegraphics[height=0.28\linewidth]{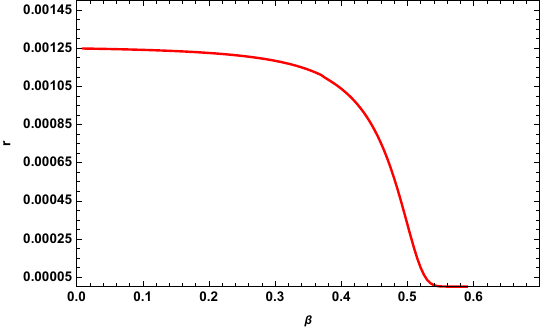}}
    \caption{The observations $n_s$ \textit{vs} $r$ (a) and the second-order derivative of potential $V_{xx}(x_{M_2})$ (b) for the models with given $a=1/3.0001$ (blue-dashed line), $a=1/3.001$ (red-solid line), and $a=1/3.01$ (green-dot-dashed line). The dependence of $n_s$ (c), $r$ (d) on the parameter $\beta$ for the model with $a=1/3.001$ by setting $N=55$.}
    \label{fig:t2}
\end{figure}

Setting $\beta=1$, the potential is
\begin{equation}
    V(x)= \tanh ^2\left(\frac{x}{\sqrt{6a} }\right)\left(\tanh \left(\frac{x}{\sqrt{6a} }\right)-1\right)^2 \text{sech}^2\left(\frac{x}{\sqrt{6a} }\right)^{1-3 a }~.
\end{equation}
As $a$ approaches $1/3$, the potential undergoes further simplification, resulting in $V\simeq y^2(1-y)^2$ where $y=\tanh{x/\sqrt{2}}$. This simplified form closely resembles the supersymmetric models discussed in \cite{Li:2014zfa}, which feature a potential of the form $V=|a_1\phi-a_2\phi^2|^2$ with $a_1,a_2$ being positive real numbers. However, these models predict a larger value of $r$ than the latest CMB constraints. Conversely,  the inflationary models embedded in no-scale supergravity exhibit a value of $r\sim 1\times10^{-3}$. Setting $N=55$, the numerical results for scalar spectral index $n_s$ \textit{vs} the tensor-to-scalar ratio $r$, second-order derivative of potential $V_{xx}$ for the given value of $a$, as well as the dependence of predictions on $\beta$ are presented in Fig. \ref{fig:t2}.

As illustrated in Fig. \ref{fig:t2}(b), the derivative of the potential, $-V_{xx}$, exceeds the value of $0.1$ for models with the parameter $\beta>1$. 
Consequently, the second slow-roll parameter, $\eta$, falls much below $-0.3$, resulting in the corresponding spectral index, $n_s$, falling outside the Planck/BICEP range. This observation is evident from Fig. \ref{fig:t2}(c). Therefore, the trajectory from $x_{M_2}$ to $x_{m_1}(=0)$ is prohibited. Similarly, the inflationary trajectory from $x_{M_2}$ to $x_{m_2}(>x_{M_2})$  follows a comparable pattern.
Motivated by the dependence of $V_{xx}$ on the parameter $\beta$, when exploring the viable parameter space for $\beta$ with the condition that $a$ is greater than $1/3$, for a given value of $a$, it is imperative to identify a suitable $\beta$ that ensures that the absolute value of the second-order derivative of the potential remains below $10^{-2}$.

\subsubsection{Trajectory III: $x_{m_2}<x_e<x_i<x_{M_3}$}
\begin{figure}[!h]
    \centering
    \subfigure[]{\includegraphics[height=0.27\linewidth]{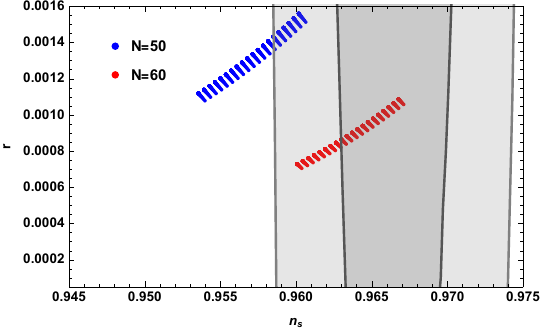}}\quad
    \subfigure[]{\includegraphics[height=0.28\linewidth]{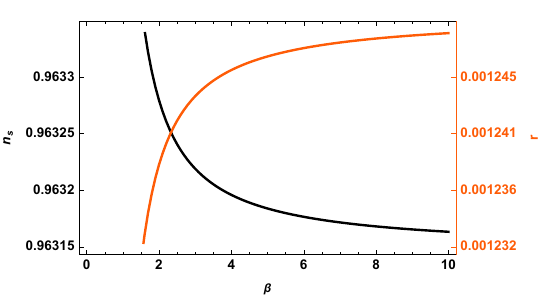}}
    \caption{(a) The observations $n_s$ vs $r$ for the models with trajectory III. (b) The dependence of $n_s$ and $r$ on the parameter $\beta$ for the model with given $ a=1/3.001$ and setting $N=55$. }
    \label{fig:t3}
\end{figure}

The last allowable trajectory is from the local maximum $x_{M_3}$ to the local minimum $x_{m_2}$. The numerical results for scalar spectral index $n_s$ \textit{vs} the tensor-to-scalar ratio $r$ and the dependency of predictions on $\beta$ are presented in Fig. \ref{fig:t3}. In contrast to models on the previous trajectories, $n_s$ or $r$ decreases or increases monotonically with $\beta$.

\subsection{$a=1/3$: supersymmetric model with $V=|a_1\phi-a_2\phi^2|^2$}

\begin{figure}[h]
    \centering
    \subfigure[]{\includegraphics[height=0.28\linewidth]{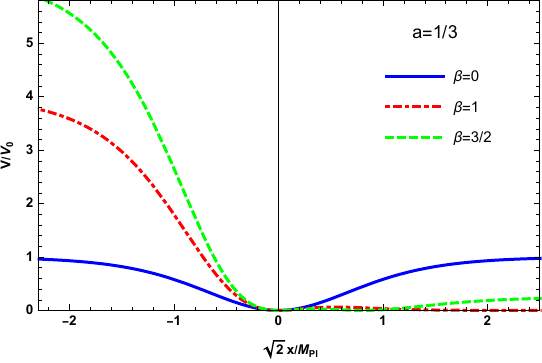}}\quad
    \subfigure[]{\includegraphics[height=0.28\linewidth]{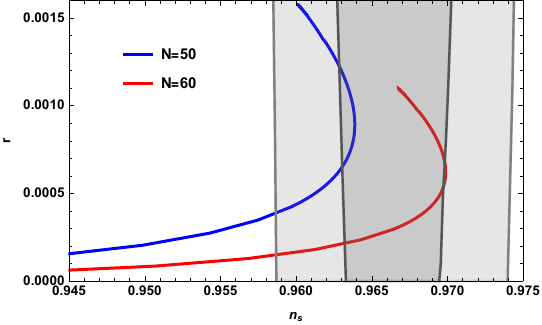}}
    \caption{Potential (a) and the predictions $n_s$ vs $r$ (b) for the models with $a=1/3$.}
    \label{fig:V_13}
\end{figure}

The potential becomes
\begin{equation}
    V(x)=\tanh ^2\left(\frac{x}{\sqrt{2}}\right) \left(1-\beta  \tanh \left(\frac{x}{\sqrt{2}}\right)\right)^2~.\label{eq:V_13}
\end{equation}
In the limit $x\to\pm\infty$, the potential approaches $(-1\pm\beta )^2$, creating a favourable platform for inflation. The plots for the potential with varying $\beta$ are presented in Fig.~\ref{fig:V_13}. For models with $0\leq\beta\leq1/2$, a minimum is located at $x_m=0$. Models with $1/2<\beta\leq1$ exhibit a minimum at $x_m=0$ and a maximum located at $x_M=\sqrt{2} \tanh ^{-1}\left(\frac{1}{2 \beta }\right)$.  Finally, for models with $\beta>1$, there are two minima at $x_{m_{1,2}}=0,\sqrt{2} \tanh ^{-1}\left(\frac{1}{\beta}\right)$ and a maximum at $x_M=\sqrt{2} \tanh ^{-1}\left(\frac{1}{2 \beta }\right)$. Additionally, the predictions $n_s$ vs $r$ for the models are also shown in Fig.~\ref{fig:V_13}.

\begin{figure}[h]
    \centering
    \subfigure[]{\includegraphics[height=0.26\linewidth]{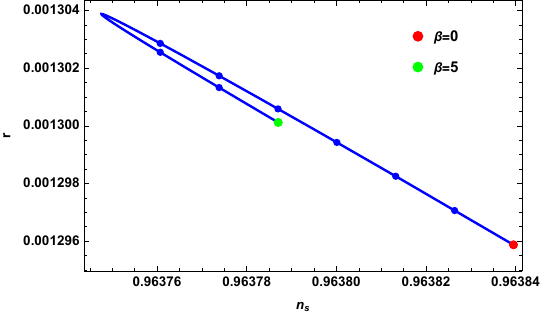}}\quad
    \subfigure[]{\includegraphics[height=0.28\linewidth]{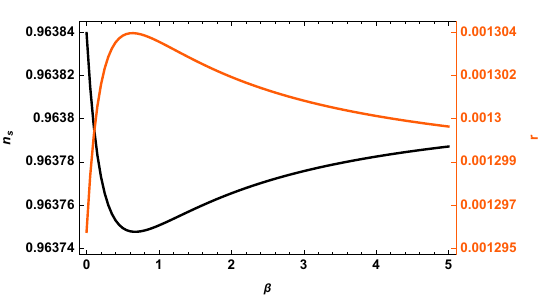}}
    \caption{The observations $n_s$ \textit{vs} $r$ for the models with potential \eqref{eq:V_13} and the trajectory is on the regime $x<0$, consistent with the $1\sigma$ region of Planck data by setting $N=55$. }
    \label{fig:nsr_13_beta}
\end{figure}
The first slow-roll parameter is
\begin{equation}
    \varepsilon(\beta,z)=\frac{\text{csch}^2(z) \text{sech}^2(z) (\coth (z)-2 \beta )^2}{(\beta -\coth (z))^2}
\end{equation}
with $z=x/\sqrt{2}$. The tensor-to-scalar ratio $r=16\varepsilon$ depends on the initial inflation point $x_i(z_i)$ and the parameter $\beta$. By fixing $N=55$, the dependency of $n_s$ and $r$ on $\beta$ is depicted in Fig. \ref{fig:nsr_13_beta}. As the parameter $\beta$ increases, the spectrum index $n_s$ initially decreases and then increases, whereas the tensor-to-scalar ratio $r$ first rises and subsequently falls.  Eventually, both settle at a particular value. 

In the limit $\beta$ approaches infinity and $z_i$ is approximately $-3.4~M_{\rm Pl}$, the slow-roll parameters are,
\begin{equation}
    \begin{split}
        \varepsilon(z)&\simeq 16 \text{csch}^2\left(2z\right)\\
        &=\frac{64}{(e^{2z}-e^{-2z})^2}
        \simeq64e^{4z}~,\\ 
        \eta(z)&\simeq -8 \left(\cosh \left(2z\right)-4\right) \text{csch}^2\left(2z\right)\\
        &=-\frac{16 e^{2 z} \left(1-8 e^{2 z}+e^{4 z}\right)}{\left(1-e^{4 z}\right)^2}
        \simeq-16e^{2z}~.
    \end{split}
\end{equation} 
Then, the CMB predictions are
\begin{equation}
    \begin{split}
        n_s&=1+2\eta-6\varepsilon \simeq 1-32e^{2z}-384e^{4z}\sim 0.9638~,\\
        r&=16\varepsilon \simeq 1024e^{4z} \sim 1.27\times 10^{-3}~.
    \end{split}
\end{equation}
They are consistent with the constraints from Planck and BICEP group, and will be tested by the following QUBIC experiments.

\subsection{$1/3<a<1$}

\begin{figure}[t]
    \centering
    \subfigure[]{\includegraphics[height=0.26\linewidth]{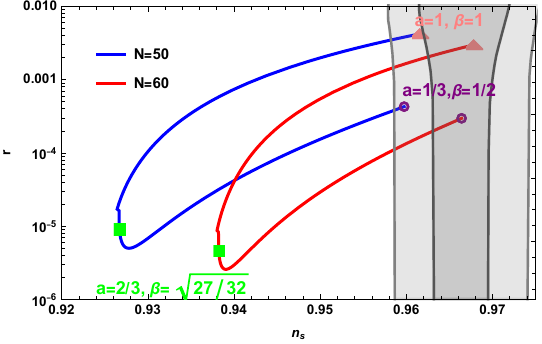}}\quad
    \subfigure[]{\includegraphics[height=0.26\linewidth]{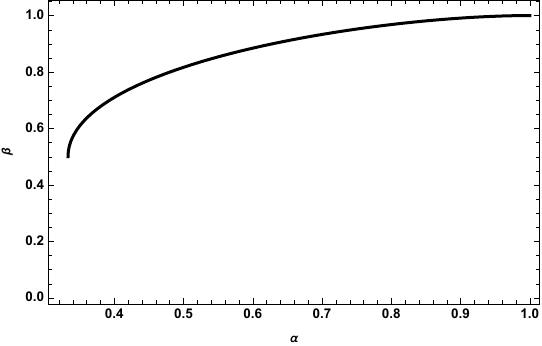}}
    \caption{The observations $n_s$ \textit{vs} $r$ (a) for the models with parameter space $P_1$ (b). }
    \label{fig:nsr_para}
\end{figure}

\begin{figure}[t]
    \centering
    {\includegraphics[height=0.3\linewidth]{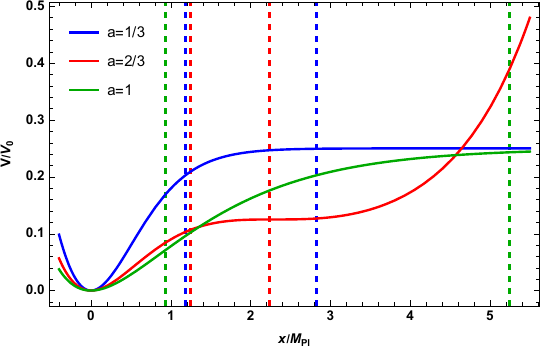}}
    \caption{ The potential for the models with parameter sets $(a,~\beta_1)=(1/3,~1/2)$, $(a,~\beta_1)=(2/3,~\sqrt{27/32})$, and $(a,~\beta_1)=(1,~1)$ respectively. The vertical lines are the corresponding inflationary initial and end points.}
    \label{fig:V_Vxx}
\end{figure}

Inspired by the dependence of the predictions $n_s$ and $r$ on the parameter $\beta$, we identify parameter space $P_1$ for $a-\beta$ such that the potential fulfills the relations  
\begin{equation}
    V_x(a,\beta,x)= 0~,~~ V_{xx}(a,\beta,x)=0~, ~~\text{with~} x>0.\label{eq:Vx_Vxx}
\end{equation}
In this manner, the inflationary plateaus emerge naturally, and the predicted observations $n_s$ and $r$ are depicted in Fig. \ref{fig:nsr_para}. The circles, squares, and triangles correspond to the parameter sets $(a,~\beta_1)=(1/3,~1/2)$, $(a,~\beta_1)=(2/3,~\sqrt{27/32})$, and $(a,~\beta_1)=(1,~1)$ respectively. Here, for a given $a$, we denote the corresponding $\beta$ that satisfies the relations in Eq. \eqref{eq:Vx_Vxx} as $\beta_1$. 
Additionally, Fig. \ref{fig:V_Vxx} displays the corresponding potential and the initial/end points for the inflaton $x$. 
The non-smooth segment in the figure is attributed to the different inflation-end condition. Specifically, inflation terminates at $\varepsilon=1$ when the model parameter satisfies $a<91/120$. Conversely, for $a>91/120$ inflation ends at $\eta=-1$. This phenomenon is also evident in the $\alpha$-attractor E-Model~\cite{Lin:2023jls}, where slow-roll inflation ends at $\varepsilon=1$ for models with $a>1/3$ and at $\eta=-1$ for models with $a<1/3$. 
For the models within the parameter space $P_1$, most predictions of $n_s$ fall below the observable values. With $N=60$, the favourable parameter space in $P_1$ is narrow, specifically $1/3\leq a\leq10027/30000$ as well as $9933/10000\leq a\leq 1$.


\begin{figure}[t]
    \centering
    {\includegraphics[height=0.3\linewidth]{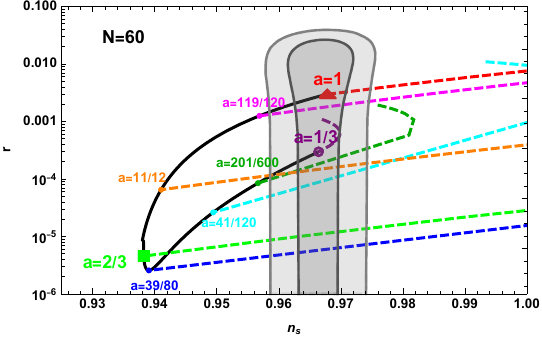}}
    \caption{The observations $n_s$ \textit{vs} $r$ for the models with parameter $1/3<a<1$.}
    \label{fig:nsr-all}
\end{figure}

Actually, to extend the range of our predictions, it is necessary to establish the parameter space $P_2$ for $a-\beta$ to ensure that the potential satisfies the relations,
\begin{equation}
    V_x(a,\beta,x)\simeq 0~,~~V_{xx}(a,\beta,x)=0~,~~ \text{with~} x>0.
\end{equation}
We find that a model with $\beta_2$ slightly smaller than $\beta_1$ is capable of yielding cosmological predictions that encompass the entire Planck/BICEP plane. The results are shown in Fig.~\ref{fig:nsr-all}, featuring eight representative classes of benchmark points for models with $a=1/3,~201/600,~41/120,~39/40,~2/3,~11/12,~119/120$ and $1$. Notable, the models with $1/3<a\leq1$ predict the tensor-to-scalar ratio $r<10^{-3}$, which is compatible with current and future CMB measurements.

\section{Formation of primordial black holes and scalar-induced gravitational waves}\label{sect:pbh}

Primordial black holes (PBHs) are formed from the collapse of primordial fluctuations during radiation domination. In our recent paper~\cite{Wu:2021zta}, we have demonstrated that multi-moduli models are more advantageous than single-modulus models for obtaining broad-band gravitational spectra.  Figure~\ref{fig:nsr-all} illustrates that the models with $a=39/80$ predict the smallest tensor-to-scalar ratio, specifically $r\sim 10^{-5}$. Consequently, in this section, we will delve into the formation of PBHs and SIGWs for the models with $a=39/80$. To introduce an inflection point into the potential, following the approach outlined in Ref. \cite{Wu:2021zta}, we incorporate an exponential term $-A e^{-B (\varphi^2 + \bar{\varphi}^2)} (\varphi^2 + \bar{\varphi}^2)$  into the K\"ahler potential. Near this inflection point, the potential exhibits a perfect flat plateau, where the slow-roll condition $\eta\ll 1$ is violated, and inflation undergoes an ultra-slow-roll phase. As a result, the power spectrum is significantly enhanced to $\mathcal{O}(10^{-2})$ on small scales. This enhancement will induce the second-order tensor perturbations after the horizon reentry, stemming from the nonlinear coupling between tensor perturbations and scalar perturbations.

\begin{table}[t]
    \centering
    \begin{tabular*}{0.85\textwidth}{@{\extracolsep{\fill}}c|ccccccc}
    \hline
         ~Model &  $\beta$  & $A$ & $B$ &$x_*/M_{\rm Pl}$& $n_s$ & $r$ & $N$~ \\ \hline
         $M_1$ & 0.8181 & 0.4564032 & 5.4 & 2.034 & 0.9685 & $9.1\times10^{-5}$ & 51.6 \\ 
         $M_2$ & 0.8185 & 0.4528304 & 5.3 & 2.035 & 0.9642 & $1.4\times10^{-4}$ & 52.7 \\ 
         $M_3$ & 0.818  & 0.4497767 & 5.2 & 2.041 & 0.9689 & $5.6\times10^{-4}$ & 52.4 \\ 
         $M_4$ & 0.814 & 0.447051 & 5.05 & 2.050 & 0.9737 & $5.3\times10^{-3}$ & 56.5 \\ 
         \hline
    \end{tabular*}
    \caption{The parameters for the model with $a=39/80$ and $c_1=1$.}
    \label{tab:model_para}
\end{table}
\begin{table}[t]
    \centering
    \begin{tabular*}{0.85\textwidth}{@{\extracolsep{\fill}}c|ccccccc}
    \hline
         ~Model & $k/{\rm Mpc}^{-1}$ & $\mathcal{P_R}$ &$M_{PBH}/M_{\odot}$ & $f_{PBH}$ & $f_{GW}/{\rm Hz}$ ~\\ \hline
         $M_1$ & $9.6\times10^{13}$ & 0.023 & $4.0\times10^{-16}$ & 0.41 & 1.1\\ 
         $M_2$ & $2.3\times10^{12}$ & 0.025 & $7.2\times 10^{-13}$ & 0.39  & 0.026\\ 
         $M_3$ & $1.2\times10^{9}$ & 0.031 & $2.5\times10^{-6}$ & 0.096 & $1.2\times10^{-5}$  \\
         $M_4$ & $1.7\times10^{5}$ & $3.3\times10^{-3}$ & - & - & $1.6\times10^{-9}$ \\ 
    \hline
    \end{tabular*}
    \caption{The peak values of the power spectrum, the mass and abundance of PBHs, and the frequency of SIGWs.}
    \label{tab:pbh_sigw}
\end{table}

Here, we present four benchmark points in Table \ref{tab:model_para}. The parameters are set as $A\sim 0.5$ and $B\sim 5$ to ensure that the additional term can be disregarded, thereby preserving the overall slow-roll story with the exception of introducing an inflection point. The results for $n_s$ and $r$ at horizon exit are consistent with the CMB constraints $n_s=0.9649\pm 0.0042$ and $r\leq 0.032$ \cite{Planck:2018jri,BICEP:2021xfz}.
The corresponding peak of power spectrum, the mass and abundance of formed PBHs and the frequency of SIGWs are listed in Table \ref{tab:pbh_sigw}.  When the primordial fluctuations are assumed to be Gaussian, the abundance of PBHs is proportional to $f_{PBH}\propto \sqrt{P_{\mathcal{R}}} \exp{(-P_{\mathcal{R}}^{-1})}$, exhibiting exponential sensitivity to the amplitude of primordial fluctuations. Thus, to determine the composition of dark matter in the universe, the fine-tuning of inflationary parameter $a$ is indeed necessary~\cite{Azhar:2018lzd,Nakama:2018utx,Cole:2023wyx}.

\begin{figure}[t]
    \centering
    \includegraphics[height=0.35\linewidth]{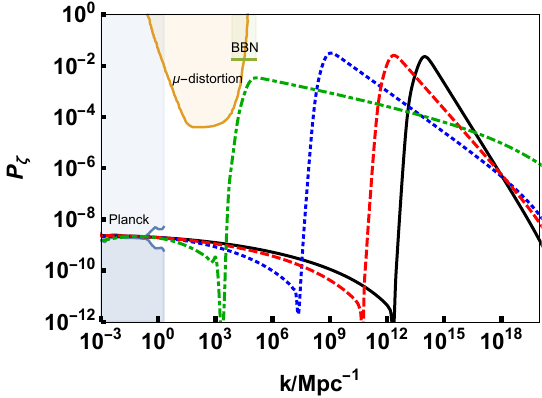}
    \caption{The enhanced scalar power spectrum. The black solid line, red dashed line, blue dotted line, and green dot-dashed line correspond to the Model $M_1$-$M_4$, respectively. See text for constraint details.}
    \label{fig:pow_spectrum}
\end{figure}

\begin{figure}[t]
    \centering
    \subfigure[]{\includegraphics[height=0.32\linewidth]{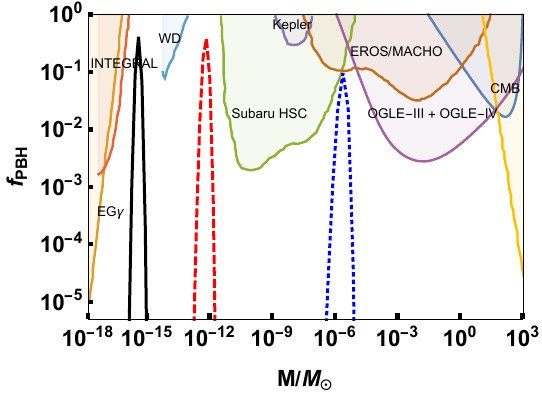}}\quad
    \subfigure[]{\includegraphics[height=0.32\linewidth]{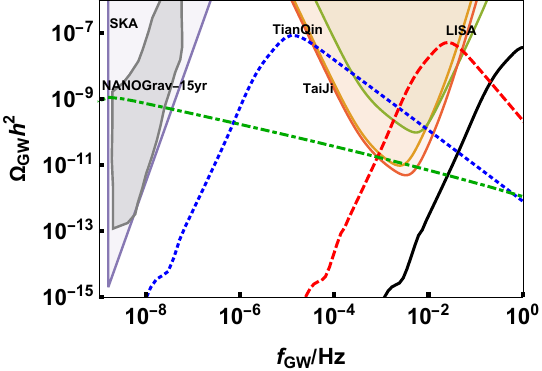}}
    \caption{The PBH abundances (a) and the energy densities of SIGWs (b). The black solid line, red dashed line, blue dotted line,  and green dot-dashed line corresponds to the Model $M_1$-$M_4$, respectively. See text for constraint details.}
    \label{fig:pbh_sigw}
\end{figure}

We plot the numerical evolution for power spectrum in Fig. \ref{fig:pow_spectrum}, the corresponding mass/abundance of PBHs and energy density of SIWGs in Fig. \ref{fig:pbh_sigw}. The constraints in these figures are from Refs. \cite{Fixsen:1996nj,Danzmann:1997hm,Tisserand:2006zx,Carr:2009jm,Griest:2013esa,Moore:2014lga,Graham:2015apa,TianQin:2015yph,Inomata:2016rbd,Ali-Haimoud:2016mbv,Poulin:2017bwe,Niikura:2017zjd,Raidal:2017mfl,Ali-Haimoud:2017rtz,Cai:2019amo,Hu:2017mde,NANOGrav:2023gor,Mroz:2024mse,Mroz:2024wag}. The results in figures are distinguished by black solid lines ($M_1$), red dashed lines ($M_2$), blue dotted lines ($M_3$), and green dot-dashed lines ($M_4$). The formed PBHs with masses of $\mathcal{O}(10^{-16}M_{\odot})$ and $\mathcal{O}(10^{-12}M_{\odot})$ can make up all of dark matter, with the abundances approaching $f_{PBH}\simeq1$. On the other hand, the PBH with mass of $\mathcal{O}(10^{-6}M_{\odot})$ only contribute partially to dark matter with $f_{PBH}\simeq 0.1$. 
In model $M_4$, a broad-band stochastic gravitational wave background is generated, accounting for a potential source for the NANOGrav 15-year signal \cite{NANOGrav:2023gor}. The SIWGs generated in models $M_1$ and $M_2$ will be probed by the space-based GW detector, such as LISA \cite{Danzmann:1997hm}, Taiji \cite{Hu:2017mde} and TianQin \cite{TianQin:2015yph}.
On the other hand, the SIWGs generated in model $M_3$ will be tested by the
future ground-based GW observatories.

\section{Conclusion}\label{sect:con}
The generic no-scale inflationary models are characterized by a real parameter $0<a\leq 1$ that surveys the cosmological predictions for these models, namely the scalar spectral index and the tensor-to-scalar ratio, as well as the production of primordial black holes and scalar induced gravitational waves. Under the slow-roll conditions, we have systematically investigated the potential inflationary trajectory and subsequently calculated, both analytically and numerically, the spectrum index $n_s$ and tensor-to-scalar ratio $r$. 
The obtained small value of $r$ is not only consistent with the current observations from Planck/BICEP/Keck Array group, but it is also likely to hold true under future tighter constraints. Notably, the predicted $n_s~ vs~ r$ in these models covers the whole observed plane. Furthermore, the tensor-to-scalar ratio $r$ can be smaller than $10^{-5}$, making these model particularly suitable for discussing the generation of primordial black holes and scalar-induce gravitational waves.   

By incorporating an exponential term into the K\"ahler potential, an inflection point is introduced into the scalar potential.  This addition does not affect the original slow-roll inflation, except for the insertion of an ultra-slow-roll phase. Thus, the power spectrum for the primordial curvature perturbation is significantly enhanced to be $\mathcal{O}(0.1)$. When the scalar perturbations and tensor perturbations couple at the nonlinear level, large primordial curvature perturbations at small scales induce second-order tensor perturbations. 
Subsequently, primordial black holes with a wide range of masses are formed, potentially accounting for the entirety or a portion of dark matter. The production of broad-band stochastic gravitational waves is compatible with the observations made by pulsar time array experiments, such as NANOGrav. On the other hand, peak-like gravitational waves will be tested by the upcoming space-based or ground-based gravitational wave observatories.

\acknowledgments

L Wu is supported in part by the Natural Science Basic Research Program of Shaanxi, Grant No. 2024JC-YBMS-039 and No. 2024JC-YBMS-521. T Li is supported in part by the National Key Research and Development Program of China Grant No. 2020YFC2201504, by the Projects No. 11875062, No. 11947302, No. 12047503, and No. 12275333 supported by the National Natural Science Foundation of China, by the Key Research Program of the Chinese Academy of Sciences, Grant No. XDPB15, by the Scientific Instrument Developing Project of the Chinese Academy of Sciences, Grant No. YJKYYQ20190049, and by the International Partnership Program of Chinese Academy of Sciences for Grand Challenges, Grant No. 112311KYSB20210012.



 \bibliographystyle{JHEP}
 \bibliography{main.bib}


\end{document}